\documentclass[aps,prb,twocolumn,showpacs]{revtex4-1}
\usepackage{graphicx}
\usepackage{hyperref}
\begin{document}

\title{Metastability effects in strained and stressed SrTiO$_3$ films}

\author{Alexander I. Lebedev}
\email[]{swan@scon155.phys.msu.ru}
\affiliation{Physics Department, Moscow State University, 119991 Moscow, Russia}

\date{\today}

\begin{abstract}
The sequence of ground states for SrTiO$_3$ film subjected to epitaxial strain
as well as to mechanical stress along the [001] and [110] axes is calculated
from first principles within the density functional theory. Under the
fixed-strain boundary conditions, an increase in the lattice parameter of a
substrate results in the $I4cm \to I4/mcm \to Ima2 \to Cm \to Fmm2 \to Ima2$(II)
sequence of ground states. Under the fixed-stress boundary conditions, the phase
sequence is different and depends on how the stress is applied. It is revealed
that the simultaneous presence of competing ferroelectric and antiferrodistortive
instabilities in SrTiO$_3$ gives rise to the appearance of metastable phases,
whose number increases dramatically under the fixed-stress conditions. In the
metastable phases, the octahedral rotation patterns are shown to differ
substantially from those in the ground state. It is suggested that in systems
with competing instabilities, each polar phase has its optimal octahedral
rotation pattern which stabilizes this phase and creates a potential barrier
preventing this phase to be transformed into other structures.
\end{abstract}


\maketitle

\section{Introduction}

Elastic strain is widely used today to improve the properties of electronic
materials. For example, the effect of the strain-induced decrease of the
acceptor binding energy in germanium enabled to create unique photodetectors
working in the far infrared.~\cite{InfraredDetectors}  An increase of the electron
mobility in strained silicon enabled to considerably improve the performance of
silicon field-effect transistors,~\cite{StrainedSilicon}  and the use of highly
strained layers in pseudomorphic high electron mobility transistors made it
possible to deliberately
tune the energy diagrams of these heterostructures in order to significantly
improve their characteristics.~\cite{pHEMTTechnolAppl}

The appearance of a strain-induced ferroelectricity in different dielectrics
significantly extends the functionality of these materials and enables to offer
new, previously known designs of electronic devices. The strain of incipient
ferroelectrics leads to particularly impressive results.~\cite{PhysRevB.61.R825,
Nature.430.758,PhysRevLett.97.267602,PhysStatSolidiB.244.3660,
ApplPhysLett.90.242918,Science.324.367,PhysRevB.87.134102}  For example, the
stretching of thin films of strontium titanate SrTiO$_3$ grown on DyScO$_3$
substrates increases the Curie temperature in this material, which is nonpolar
in the absence of strain, to $\sim$300~K,~\cite{Nature.430.758}  and the
compression of these films grown on silicon substrates increases this temperature
up to $\sim$410~K (Ref.~\onlinecite{Science.324.367}).

Strontium titanate is an incipient ferroelectric exhibiting a competition between
the ferroelectric and antiferrodistortive (octahedral rotational)
instabilities.~\cite{PhysRevLett.74.2587} The latter of them is the cause of
the phase transition observed in SrTiO$_3$ at about 105~K. The first
work in which the influence of strain on the ferroelectric properties of
strontium titanate was considered within the phenomenological approach was the
work of Uwe and Sakudo.~\cite{PhysRevB.13.271}
Pertsev \emph{et al.}~\cite{PhysRevB.61.R825} have developed this approach to
describe the effect of the epitaxial strain on SrTiO$_3$ thin films. By expanding
the thermodynamic
potential in a power series of two order parameters (polarization and octahedral
rotation) to the fourth order, the authors obtained a rich pressure--temperature
phase diagram with a
large number of different phases. For the set of material constants of SrTiO$_3$
used by the authors, all obtained solutions had the order parameters directed
along the axes of the cubic structure. Subsequent first-principles calculations
in which the antiferrodistortive instability was neglected,~\cite{PhysRevB.71.024102,
PhysRevB.72.144101,JapJApplPhys.44.7134}  however, have shown that in stretched
SrTiO$_3$ films the polarization should be directed along the [110] axis, in
agreement with the experimental data.~\cite{PhysRevLett.97.257602,PhysRevB.73.184112}
A more thorough first-principles study of the low-temperature phases of SrTiO$_3$,
in which both ferroelectric and antiferrodistortive instabilities were taken
into account,~\cite{JApplPhys.100.084104}  showed that in highly stretched
SrTiO$_3$ films the polarization is indeed directed along the [110] axis, but at
low strain a phase with the [100] polarization appears. This orientation of the
polarization was observed later in anisotropically strained SrTiO$_3$ films
grown on GdScO$_3$ and DyScO$_3$ substrates.~\cite{ApplPhysLett.92.192902,PhysRevB.79.224117}
The phase diagrams of strained SrTiO$_3$ for other sets of material constants
were studied within the phenomenological and the phase-field approaches in
Refs.~\onlinecite{PhysRevB.73.184112,PhysSolidState.51.1025,ApplPhysLett.96.232902,
JApplPhys.108.084113}.

When studying the strain effects on the properties of different materials, the
fixed-strain boundary conditions are usually used. However, the study of the
fixed-stress boundary conditions is also important for real systems. To explain
this, let's consider a thin epitaxial film which has domain structure and is
fixed on a substrate. If the substrate is incompressible, then the in-plane
lattice parameters in these domains would be equal to the lattice parameter of
the substrate. However, in real systems the lattice parameters in the domains
will be different because the substrate is compressible, and the difference
between them will increase with increasing film thickness. In the limit of highly
compressible (flexible) substrate, if one neglects mechanical and electrical
boundary conditions at the domain walls, we come to the fixed-stress
boundary conditions. In practice, the fixed-strain boundary conditions can be
realized in ultrathin epitaxial films, whereas the fixed biaxial stress boundary
conditions can be realized in films grown on flexible substrates or thin plates
which are fixed at their edges on a set of piezoelectric actuators. The actual
boundary conditions in real systems are intermediate between the two limiting
cases, and this is why the analysis of both cases is necessary when studying the
strain effects.

In this work, we consider the ferroelectric and antiferrodistortive instabilities
in SrTiO$_3$ under the fixed-stress boundary conditions and compare them with
the results obtained for the fixed-strain boundary conditions. Two different
ways of applying stress to the film are examined: a uniaxial stress normal to
the film plane and a biaxial stress in the film plane. We show that these two
ways of applying stress result in different phase diagrams. We demonstrate that
the simultaneous presence of competing ferroelectric and antiferrodistortive
instabilities
in SrTiO$_3$ gives rise to the appearance of previously unknown metastable
phases, whose number increases dramatically under the fixed-stress conditions,
and explain the origin of the metastability.

\section{Calculation technique}

Elastic deformation of SrTiO$_3$ films was realized in two ways. To create the
fixed-strain boundary conditions, the (001)-oriented film was grown on a cubic
substrate with the lattice parameter $a_0$, which was varied within $\pm$2\% of
the lattice parameter of cubic strontium titanate. To create the fixed-stress
boundary conditions, the stress was applied along the $z$~axis normal to the
film plane (the pressure is $p_{[001]}$) or in the $xy$ plane (the pressure is
$p_{[110]}$). The film was considered free to relax in directions normal to
the applied stress or strain.

Calculations of the equilibrium lattice parameters and atomic positions in films
were performed within the first-principles density functional theory using the
\texttt{ABINIT} software. The exchange-correlation interaction was described in
the local density approximation (LDA). The pseudopotentials of atoms constructed
using the RKKJ scheme~\cite{PhysRevB.41.1227}  were borrowed from
Ref.~\onlinecite{PhysSolidState.51.362}. The plane-wave cutoff energy was 30~Ha
(816~eV), the accuracy of self-consistent energy calculations was better than
$10^{-10}$~Ha. For the integration over the Brillouin zone, a 8$\times$8$\times$8
Monkhorst-Pack mesh for the cubic cell or meshes with equivalent density of
$k$-points for low-symmetry phases were used. The relaxation of atomic positions
and lattice parameters was performed until the forces acting on the atoms become
less than $2 \cdot 10^{-6}$~Ha/Bohr (0.1~meV/{\AA}). Near the boundaries between
the phases, the accuracy of the relaxation was increased to
$2 \cdot 10^{-7}$~Ha/Bohr.

As the energy difference between some phases in this work can be as small as
0.1~meV, the accuracy and convergence of calculations are very important issues.
Insufficient $k$-point density, parallel shift of all atoms in the unit cell,
and different sets of $k$-points for different structures are the main sources
of errors. The convergence studies have shown that an increase of the $k$-point
density by a factor of three changes the energies of phases by no more than 0.02~meV.
More important errors may result from a parallel shift of all atoms in the unit
cell. The standard deviation of the energy of phases calculated for 20~different
random parallel shifts was equal to 0.028~meV and was independent of the $k$-point
density. As to the third source of errors, in most calculations we used the
same sets of the $k$-points when performing the integration over the Brillouin
zone, but the tests have shown that the changes in the energy of phases calculated
for different sets of $k$-points do not exceed 0.03~meV. The errors in the enthalpy
calculations resulting from variations of the unit cell volume and local stresses
were less than 0.003~meV. So, the accuracy of our calculations enables to reliably
distinguish the energy difference of 0.1~meV.

In searching for the ground state, after the relaxation of a structure of each
relevant phase for each value of applied stress or strain, the phonon spectrum
and the elastic tensor were calculated and it was checked whether all optical
phonon frequencies at all high-symmetry points of the Brillouin zone are positive
and whether the determinant and all leading principal minors constructed
from components of the elastic tensor are positive (the stability criterion).
If the criterion is not satisfied, small distortions corresponding to the least
stable phonon were added to the structure, and the search for the ground state
was continued. The technique of phonon spectra and elastic tensor calculations
was similar to that described in Ref.~\onlinecite{PhysSolidState.51.362}.
We considered only monodomain states because domain walls usually have a positive
energy and so the multidomain solutions have a higher energy.

\section{Results}

The phonon spectrum calculations for the high-temperature $Pm{\bar 3}m$ phase
of SrTiO$_3$ confirmed a well-known result that there are three types of
instabilities in the phonon spectrum: the ferroelectric one associated with
the $\Gamma_{15}$ mode at the center of the Brillouin zone and two antiferrodistortive
instabilities with respect to rotations of the oxygen octahedra described by
$R_{25}$ and $M_3$ modes at the boundary of the Brillouin
zone.~\cite{Ferroelectrics.194.109,PhysSolidState.51.362}  Since the $M_3$
instability is weak and disappears when the mode at the $R$~point is condensed,
we will neglect it when searching for the ground state. The calculations showed
that in the absence of strain, the ground-state structure of strontium titanate
is the $I4/mcm$ phase.~\cite{PhysSolidState.51.362}

\subsection{Fixed-strain boundary conditions}

\begin{figure}
\centering
\includegraphics{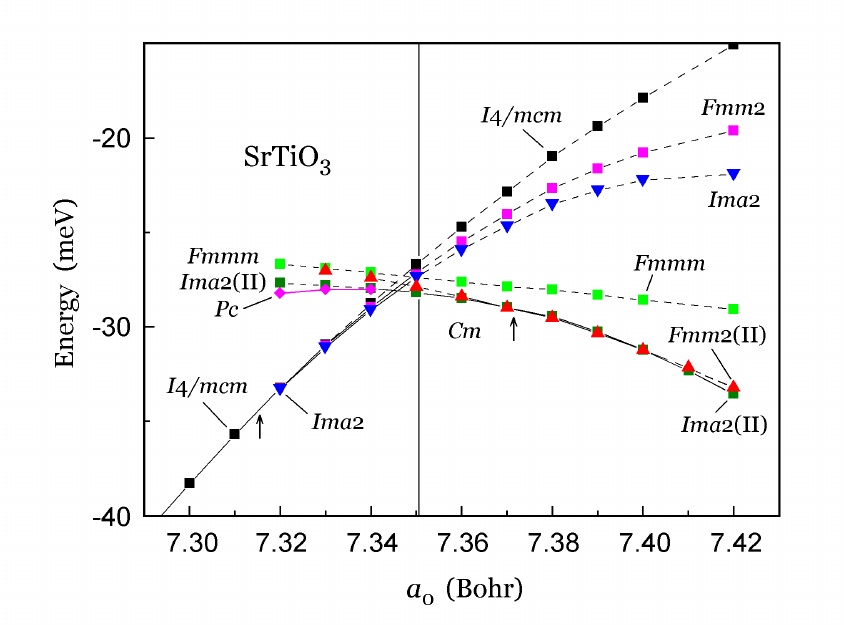}
\caption{(Color online) Energies of different phases for SrTiO$_3$ film grown on
a cubic substrate with the lattice parameter $a_0$. The energy of high-symmetry
$P4/mmm$ phase is taken as the energy reference. The vertical line indicates the
lattice parameter of cubic SrTiO$_3$. Short arrows show the lattice parameters
at which the frequencies of soft optical modes vanish.}
\label{fig1}
\end{figure}

To better understand the influence of the fixed-stress boundary conditions on
the phase diagrams, we first consider the phase diagrams for SrTiO$_3$ films grown
on a cubic substrate (the fixed-strain boundary conditions). Our calculations
(Fig.~\ref{fig1}) show that the sequence of the ground states
[$I4cm \to I4/mcm \to Ima2 \to Cm \to Fmm2$(II)${} \to Ima2$(II)] in this case is
unexpectedly complex. The transition between the $I4cm$ and $I4/mcm$ phases occurs
at $a_0 \approx 7.279$~Bohr and is not shown in the figure (see Table~A1 in the
Appendix). At $a_0 \approx 7.316$~Bohr,
the $I4/mcm$ phase becomes unstable and transforms to the $Ima2$ phase, in which
the octahedra are rotated around the [001] pseudocubic axis, whereas the
polarization is along the [110] one (Table~\ref{table1}). For all strains, this
phase has a lower energy as compared to the $Fmm2$ phase in which the rotations
are around the same axis, but the polarization is along the [100] axis. The
transition between the $Ima2$ and $Cm$ phases occurs at $a_0 \approx 7.345$~Bohr
and is a first-order transition (the molar volumes of the two phases differ by
0.75\% at the transition point). Near the phase transition point, the $Cm$ phase
can be regarded as a slightly distorted $Ima2$(II) phase,%
    \footnote{This is proven by an instability of the phonon spectra of $Ima2$(II)
    and $Fmm2$(II) phases at $a_0 = {}$7.35--7.37~Bohr. For example, in the
    $Ima2$(II) phase, the eigenvector of unstable $B_1$ phonon at the $\Gamma$~point
    at $a_0 = {}$7.36--7.37~Bohr includes, in addition to polar displacements,
    the octahedral rotations around the $y$~axis perpendicular to the polarization,
    whereas at $a_0 = 7.35$~Bohr, the unstable phonon is at the boundary of the
    Brillouin zone and describes the \emph{in-phase} octahedral rotations around
    the $z$ axis (similar to those described by the $M_3$ phonon in the cubic
    phase). The frequencies of these phonons are $\sim$10$i$ cm$^{-1}$.}
but the polarization vector quickly rotates in the plane from the [110] to [100]
direction with increasing $a_0$, and already at $a_0 = 7.38$~Bohr the $Fmm2$(II)
phase becomes the ground state. At $a_0 = {}$7.39--7.40~Bohr, the energy of the
$Fmm2$(II) phase turns out to be slightly lower than that of the $Ima2$(II) phase
(by approximately 0.1~meV, see Table~A1 in the Appendix for
the energies of the phases). It is interesting that both phases with the lowest
energy [$Fmm2$(II) and $Ima2$(II)] in this region satisfy the stability criterion
(i.e., the former of them is stable and the latter is metastable), so that in
this region the phases with polarizations along the [110] and [100] pseudocubic
axes can coexist. This, in particular, enables to explain why two different
orientations of polarization have been observed in two different
experiments~\cite{PhysRevB.73.184112,PhysRevB.79.224117} on SrTiO$_3$ films grown
on DyScO$_3$ substrates
and to understand the sensitivity of polarization to the anisotropic strain of
the substrate. At $a_0 > 7.40$~Bohr, the $Ima2$(II) phase becomes the ground state.

\begin{table}
\caption{\label{table1}Space groups and order parameters for phases considered
in this work. The first three symbols in the order parameter denote Cartesian
coordinates of the polarization vector and the last three symbols denote the
rotations around the pseudocubic axes. The lattice parameters and atomic
positions for typical representatives of the structures can be found in
Tables~A3 to A12 in the Appendix.}
\begin{ruledtabular}
\begin{tabular}{cc}
Space group & Order parameter \\
\hline
$I4/mcm$ & $(0 0 0 \, 0 0 \phi)$ \\
$I4cm$  & $(0 0 p \, 0 0 \phi)$ \\
$Fmm2$  & $(p 0 0 \, 0 0 \phi)$, $(0 p 0 \, 0 0 \phi)$ \\
$Ima2$  & $(p p 0 \, 0 0 \phi)$ \\
$Fmmm$  & $(0 0 0 \, 0 \phi 0)$, $(0 0 0 \, \phi 0 0)$ \\
$Fmm2$(II) & $(p 0 0 \, 0 \phi 0)$, $(0 p 0 \, \phi 0 0)$ \\
$Ima2$(II) & $(p p 0 \, \phi \phi 0)$ \\
$Cm$    & $(p_1 p_2 0 \, \phi_1 \phi_2 0)$ \\
$Fmm2$(III) & $(0 0 p \, 0 \phi 0)$, $(0 0 p \, \phi 0 0)$ \\
$Pc$    & $(p_1 p_1 p_3 \phi_1 \phi_1 \phi_3)$ \\
\end{tabular}
\end{ruledtabular}
\end{table}

\subsection{Fixed-stress boundary conditions}

Under the fixed-stress boundary conditions, the sequence of the ground states
resulting from compression or stretching of the film is different for two ways
of applying stress. This follows from the difference between the equations used
to calculate the enthalpy. At $p \parallel [001]$, the enthalpy is calculated
using the formula $H = E_{\rm tot} + p_{[001]} V \epsilon_3$, where $E_{\rm tot}$
is the total energy, $p_{[001]}$ is the pressure, $V$ is the unit cell volume,
and $\epsilon_3$ is the strain tensor component (in Voigt notation)
normal to the film. At $p \parallel [110]$, the enthalpy is
$H = E_{\rm tot} + p_{[110]} V (\epsilon_1 + \epsilon_2)$, where $\epsilon_1$ and
$\epsilon_2$ are the strain tensor components in the film plane.

\begin{figure}
\centering
\includegraphics{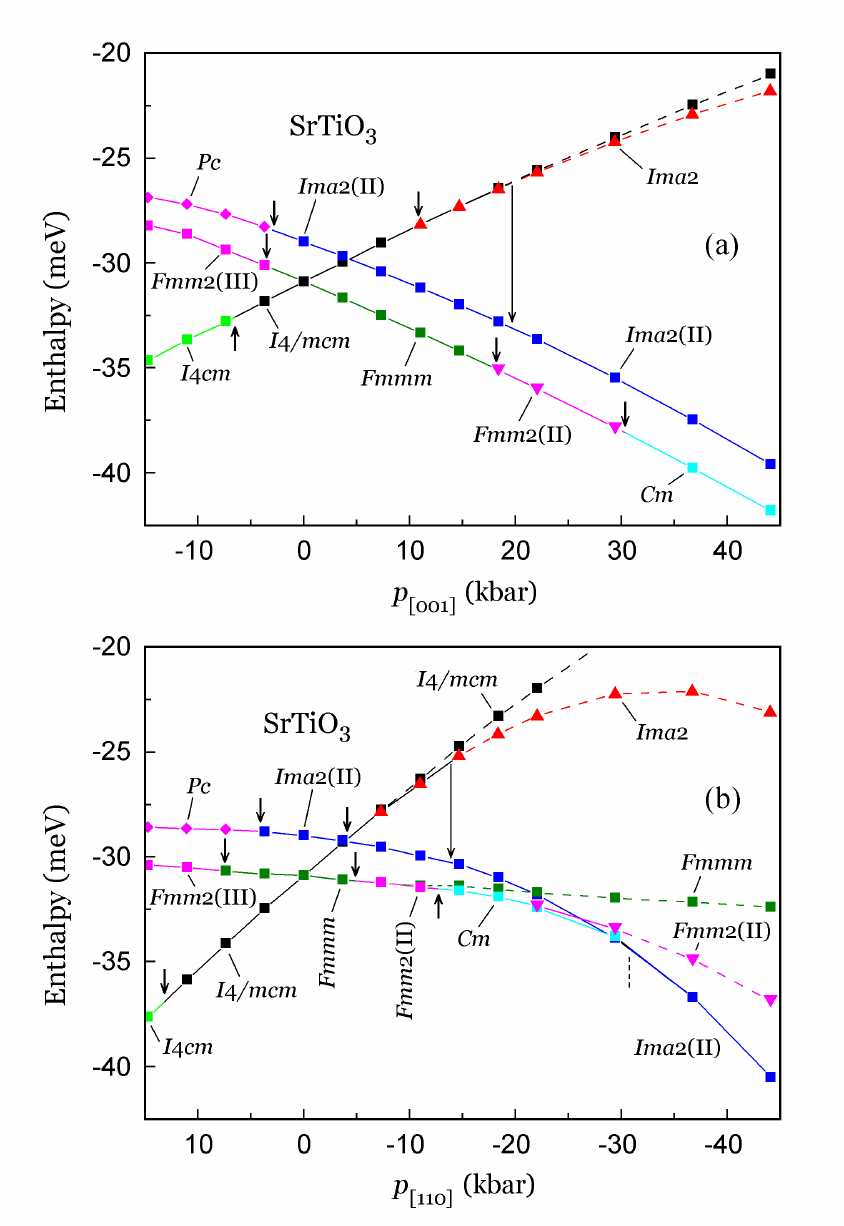}
\caption{(Color online) Enthalpies of different phases for SrTiO$_3$ film
subjected to (a) uniaxial compression along the $z$~axis and (b) biaxial
stretching in the $xy$ plane. The enthalpy of high-symmetry $P4/mmm$ phase is
taken as the energy reference. Short arrows indicate the stresses at which the
frequencies of soft phonons vanish. Solid lines connecting the points refer to
stable and metastable phases, whereas dashed lines indicate the unstable phases.
Long vertical lines with arrows show the directions of structural relaxations
at pressures where the metastable phases become unstable.}
\label{fig2}
\end{figure}

When the stress is applied, the symmetry of the high-temperature $Pm{\bar 3}m$
phase is lowered to tetragonal $P4/mmm$ phase. As for the low-temperature
$I4/mcm$ phase, the formation of two structures differing by the direction of
octahedral rotation axis is possible under stress. Because of a noticeable
spontaneous strain accompanying the octahedral rotations in the unit cell,
the resulting $I4/mcm$ phase has a lower enthalpy for biaxially compressed films
and the $Fmmm$ phase has a lower enthalpy for biaxially stretched films
(Fig.~\ref{fig2}). In this work, a positive sign of the applied stress means the
compression and a negative one means the tension. Since the compression of the
film along the $z$~axis results in its expansion in the $xy$ plane, to facilitate
the comparison of results for two ways of applying stress, we used two opposite
directions of the horizontal axis in Fig.~\ref{fig2}. The in-plane strain of the
film at a maximum pressure ($|p| = 44.1$~kbar) is 2.13\% for $p \parallel [001]$
and 1.85\% for $p \parallel [110]$.

The existence region of the $I4/mcm$ phase on the phase diagram depicted in
Fig.~\ref{fig2} is limited by the softening of the ferroelectric $A_{2u}$ mode
(under the in-plane compression) and of the ferroelectric $E_u$ mode (under
the in-plane stretching). When crossing the boundaries of the existence region
(shown by short vertical arrows in Fig.~\ref{fig2}), the frequencies of
corresponding modes become imaginary, and the symmetry of the unit cell is
lowered, respectively, to $I4cm$ and $Ima2$.%
    \footnote{Since the $E_u$ mode is doubly degenerate, to find the ground
    state resulting from the ferroelectric instability of the $I4/mcm$ phase we
    have to consider two structures described by two-component ($P$,\,0) and
    ($P$,\,$P$) order parameters and to choose among them the structure with
    a lower enthalpy. Of the two possible solutions with space groups $Fmm2$ and
    $Ima2$, the $Ima2$ phase had a lower enthalpy.}
The same applies to the $Fmmm$ phase, in which the softening of the $B_{1u}$
mode upon the in-plane compression results in the appearance of the $Fmm2$(III)
phase polarized along the $z$~axis, and the softening of the $B_{3u}$ mode upon
the in-plane stretching results in the appearance of the $Fmm2$(II) phase
polarized along the $x$~axis (we assume that in both cases the octahedral
rotations are around the $y$~axis). We note that for two ways of applying stress
we are discussing, the values at which the phase transitions occur are very
different. For example, for the $I4/mcm$--$I4cm$ phase transition the transition
pressures are $p_{[001]} = -6.5$~kbar and $p_{[110]} = +13.1$~kbar.

While there is only one $I4/mcm$--$I4cm$ phase transition in the in-plane
compressed SrTiO$_3$ films, the sequence of the ground states in stretched films
is much more complex. The calculations of the phonon spectra of the $Fmmm$ phase
show that with increasing in-plane stretching not only the polar $B_{3u}$ mode
is softened, but also the polar $B_{2u}$ mode is softened. At a pressure of
$p_{[110]} = -4.9$~kbar or $p_{[001]} = +18.2$~kbar, the former mode results
in the transition to the $Fmm2$(II) phase polarized along the $x$~axis. With
further increase of pressure (at $-$12.7~kbar or +30.3~kbar, respectively), the
$B_{2u}$ mode induces the transition to the $Cm$ phase, in which the polarization
is rotated in the $xy$ plane. For the in-plane stretching, the $Cm$ phase
transforms to the $Ima2$(II) phase at $\sim$30~kbar [see Fig~\ref{fig2}(b) and
enthalpies of different phases in Table~A2 in the Appendix]. In the case of the
out-of-plane compression, the
$Cm$ phase remains the ground state up to at least 45~kbar [Fig.~\ref{fig2}(a)].

\begin{figure}
\centering
\includegraphics{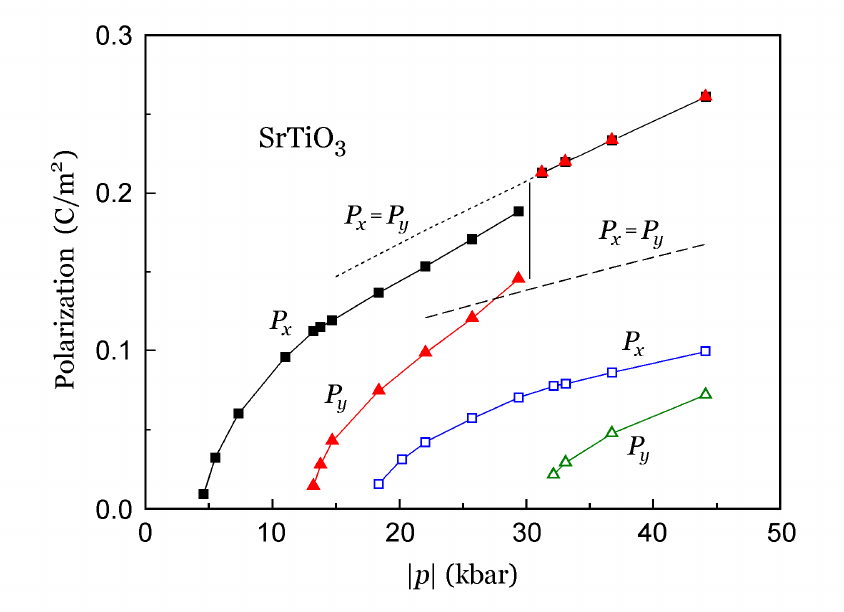}
\caption{(Color online) Components of polarization as a function of pressure for
two ways of applying stress. Data for tensile $p \parallel [110]$ stress are
shown by filled symbols and data for compressive $p \parallel [001]$ stress are
shown by open symbols. The dotted and dashed lines show the polarization in the
$Ima2$(II) phase obtained for $p \parallel [110]$ and $p \parallel [001]$,
respectively.}
\label{fig3}
\end{figure}

The pressure dependence of two components of spontaneous polarization for both
ways of applying stress is shown in Fig.~\ref{fig3}. The changes in polarization
upon the $Fmmm \to Fmm2$(II) and $Fmm2$(II)${} \to Cm$ phase transitions and
the fact that the polarization abruptly changes from the $P_x \ne P_y$ state in
the $Cm$ phase to the $P_x = P_y$ state characteristic of the $Ima2$(II) phase
are clearly seen. From the comparison of the polarizations for two ways of
applying stress it follows that the polarization is about 2.5~times higher for
biaxial stretching than for uniaxial compression, whereas the phase sequence is
the same. The reason for this is simple: the $p_{[110]} V (\epsilon_1 + \epsilon_2)$
term in the enthalpy makes larger in-plane strains more favorable under
$p_{[110]} < 0$. As the in-plane polarization strongly increases with increasing
tensile strain $(\epsilon_1 + \epsilon_2)$, the polarization in biaxially stretched
films is considerably higher than that in uniaxially compressed films. Under the
$p \parallel [001]$ compression, the polarization in $Fmm2$(II) and $Cm$ phases
is much lower than that in the metastable $Ima2$(II) phase because the rotation
around the [100] axis strongly reduces the $(\epsilon_1 + \epsilon_2)$ values.

\subsection{Hysteresis and metastability effects}

\begin{figure}
\centering
\includegraphics{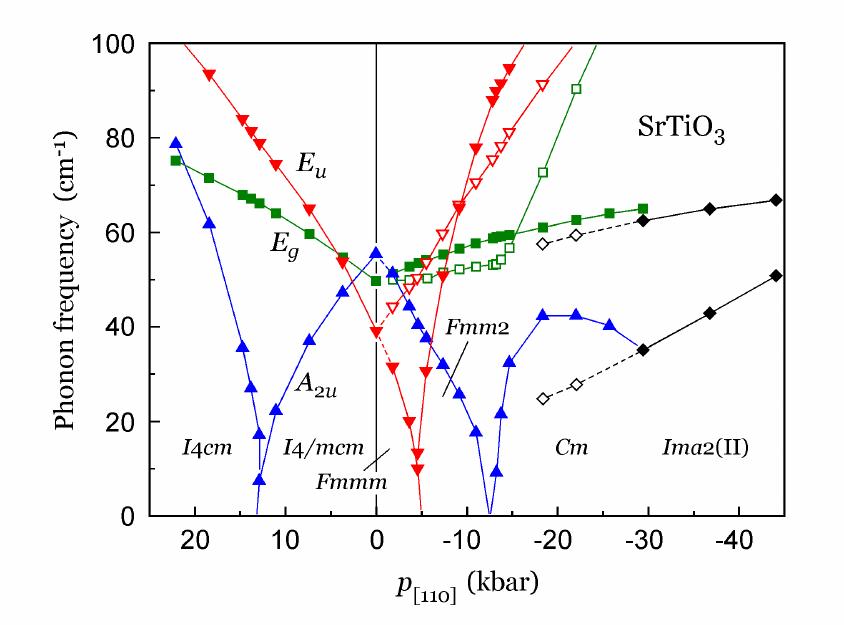}
\caption{(Color online) Frequencies of low-energy phonons in the ground-state
structures as a function of the in-plane stress ($p \parallel [110]$). Dotted
lines show the behavior of frequencies in the metastable $Ima2$(II) phase that
coexists with the ground-state $Cm$ structure.}
\label{fig4}
\end{figure}

An analysis of the stress dependence of the soft-mode frequencies in the ground
states of SrTiO$_3$ for the fixed-stress boundary conditions (Fig.~\ref{fig4})
reveals a number of previously known features. First, the pressures, at which
the extrapolated squares of the soft-mode frequencies vanish on both sides of
the $I4/mcm$--$I4cm$ phase transition, differ by 0.76~kbar for $p \parallel [110]$
(hysteresis loops are not shown in the figure).
For $p \parallel [001]$, the difference between these pressures is more
significant, 1.88~kbar. This means that in both cases the $I4/mcm$--$I4cm$ phase
transition is of the first order. Similar phenomena are observed for the
$Fmmm$--$Fmm2$(II) phase transition (the hysteresis region width is 0.48~kbar
for $p \parallel [110]$ and 0.41~kbar for $p \parallel [001]$) and for the
$Fmm2$(II)--$Cm$ phase transition (the hysteresis region width is 0.39~kbar
for $p \parallel [110]$ and 0.62~kbar for $p \parallel [001]$).

The most unusual phase transition in Fig.~\ref{fig4} is the $Cm \to Ima2$(II)
one which is observed in biaxially stretched films. As follows from
Fig.~\ref{fig3}, an abrupt change in polarization at this transition is an
evidence of the first-order phase transition. This conclusion is consistent with
the absence of any critical phonon mode softening on both sides of this
transition (Fig.~\ref{fig4}). Interestingly, that at pressures below the
transition pressure, both structures [$Cm$ and $Ima2$(II)] satisfy the stability
criterion, which means that the $Ima2$(II) phase with a higher enthalpy is
\emph{metastable} at these pressures. We attribute the transition under discussion
to a ``sliding off'' of the structure, which occurs when the enthalpy of the
$Ima2$(II) phase becomes lower than that of the $Cm$ phase: thanks to the
closeness of structures, the height of the potential barrier separating them
is low.

\begin{figure}
\centering
\includegraphics{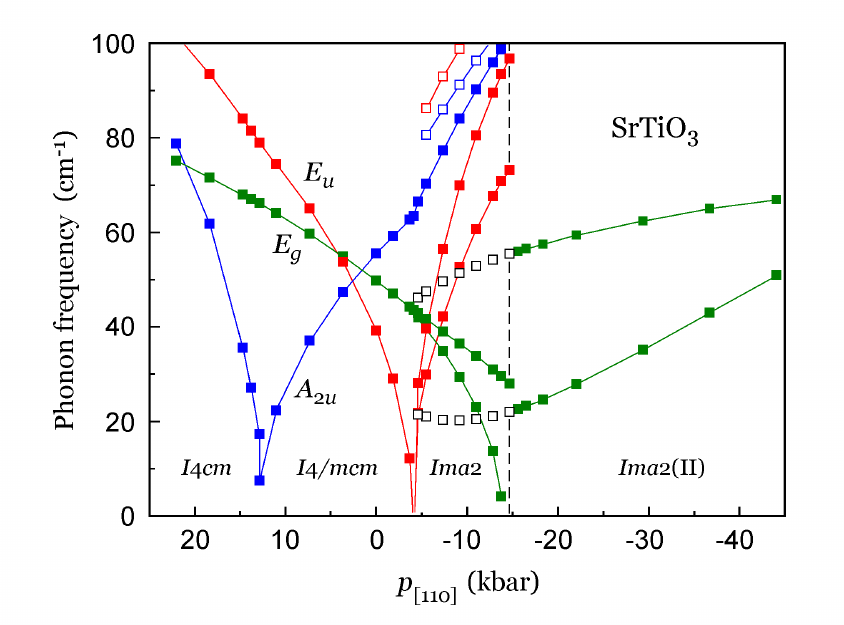}
\caption{(Color online) Frequencies of low-energy phonons in stable ($I4cm$,
$I4/mcm$) and metastable [$Ima2$, $Ima2$(II)] phases as a function of
$p \parallel [110]$ stress. The frequencies in the $Ima2$(II) phase obtained
upon releasing the in-plane stress are shown by open symbols.}
\label{fig5}
\end{figure}

It is interesting to also consider the chain of phase transitions occurring on
a branch starting from the $I4/mcm$ phase which is metastable at tensile
stresses (Fig.~\ref{fig2}).%
    \footnote{In principle, these phases can be prepared by cooling the film
    under compressive stress and after then by changing the sign of stress.}
The calculations of the phonon spectra for the $I4/mcm$ phase show that not
only the $E_u$ mode is softened with increasing tensile stress, but also the
$E_g$ mode is softened, and after the phase transition to the $Ima2$ phase at
$-$4.1~kbar it splits into two components (Fig.~\ref{fig5}). At $-$13.8~kbar,
the frequency of one of its components vanishes. Adding of the distortions
corresponding to this unstable mode to the $Ima2$ structure causes its
relaxation into a phase which has the same space group, but a different
octahedral rotations pattern. This is already familiar to us the $Ima2$(II)
phase, which was obtained earlier upon stretching the film along the
$I4/mcm \to Fmmm \to Fmm2$(II)${} \to Cm \to Ima2$(II) chain. As the $E_g$
mode originates from the $R_{25}$ mode of cubic SrTiO$_3$, the cause of the
phase transition at $-$13.8~kbar is the antiferrodistortive instability, which
results
in a change of the octahedral rotation axis direction from one normal to the
film plane in the $Ima2$ phase to one along the polar [110] pseudocubic axis
in the $Ima2$(II) phase. It is unusual here that upon the relaxation, the
$Ima2$ structure transforms through an intermediate $Cc$ phase into the
\emph{metastable} $Ima2$(II) phase, rather than into the ground-state structure
[the $Cm$ phase according to Fig.~\ref{fig2}(b)].

Similar phenomena are observed in uniaxially stressed SrTiO$_3$ films at
$p_{[001]} = +19.7$~kbar [Fig.~\ref{fig2}(a)]. The $Ima2$ structure, which
becomes unstable above this pressure, also relaxes into the metastable $Ima2$(II)
phase rather than into the ground-state structure [$Fmm2$(II) or $Cm$].

The metastability effects are also characteristic of the $Ima2$(II) phase at
$p_{[001]} = {}$15--37~kbar and at $p_{[110]} = {}-$(11--22)~kbar (Fig.~\ref{fig3}).

The behavior of the soft mode frequency in the $Ima2$(II) phase when changing
the in-plane tensile stress is also interesting. As follows from Fig.~\ref{fig5},
at the isostructural $Ima2$--$Ima2$(II) phase transition, the majority of modes
abruptly change their frequencies (dashed lines in the figure), whereas the
frequency of the soft mode does not vanish when approaching to the phase boundary
from the $Ima2$(II) phase. This means that this phase transition is of the first
order. This conclusion is also confirmed by the ability to preserve the stability
of the $Ima2$(II) phase when gradually decreasing tensile stress (open symbols
in Fig.~\ref{fig5}). Thus, in the pressure range from $-$4.1 to $-$13.8~kbar, a
hysteresis region occurs on the considered branch of the phase diagram.%
    \footnote{The stability of the metastable $Ima2$(II) phase is retained up to
    a pressure of +4.1~kbar, above which it transforms to the $Pc$ phase.}
In this region, both $Ima2$ and $Ima2$(II) phases satisfy the stability
criterion, which means, if one takes into account the enthalpies of these phases
[Fig.~\ref{fig2}(b)], that both phases are metastable. The unusual property of
the $I4/mcm \to Ima2 \to Ima2$(II) chain of transformations is that the phase
transition at $p_{[110]} = -4.1$~kbar occurs from a nonpolar to a
\emph{metastable} polar phase, which transforms into \emph{another isostructural
metastable} polar phase with increasing stress (at $-$13.8~kbar).

The properties of polar phases coexisting in the hysteresis region differ
significantly. At $p_{[110]} = -11$~kbar, the values of spontaneous polarization
are 0.128~C/m$^2$ in the $Ima2$ phase, 0.182~C/m$^2$ in the $Ima2$(II) phase,
and 0.096~C/m$^2$ in the ground-state $Fmm2$(II) phase. The band gaps calculated
in the LDA approximation are 1.795~eV in the $Fmm2$(II) phase, 1.825~eV in the
$Ima2$ phase, and 1.861~eV in the $Ima2$(II) phase.

\section{Discussion}

\subsection{Phase diagrams of strained and stressed SrTiO$_3$ films}

\subsubsection{Strained SrTiO$_3$ films}

To start, we compare the phase diagrams obtained in this work for SrTiO$_3$
film grown on a cubic substrate (the fixed-strain boundary conditions) with
the results of previous investigations.

At compressive strain, the results of this paper and previous
works~\cite{PhysRevB.61.R825,PhysRevB.73.184112,JApplPhys.100.084104} are
in good agreement and differ only in the strain values at which the
$I4/mcm$--$I4cm$ phase transition occurs. This is not surprising if we take
into account the difference in the sets of material constants used in
phenomenological models and the difference in the calculation techniques and
schemes of the pseudopotential construction used in first-principles
calculations. The difference in the $a_0$ values, at which the extrapolated
squares of the soft-mode frequencies vanish at two sides of this phase
transition, was found to be only 0.00024~Bohr in our calculations, which
indicates that the transition is close to the second-order one.

At tensile strain, the results of our and previous calculations are much more
different. In the phenomenological model,~\cite{PhysRevB.61.R825} in stretched
films at all strains the polarization is directed along the $<$100$>$ pseudocubic
axis, so that our solutions $Ima2$, $Ima2$(II), and $Cm$ are absent. In the
phenomenological model~\cite{PhysRevB.73.184112} it was shown that the $Ima2$
and $Ima2$(II) solutions appear at low values of the $\alpha_{12}$ coefficient
for the ($P^2_x P^2_y + P^2_x P^2_z + P^2_y P^2_z$) term, and the $Ima2$ phase
is stable only in a narrow region of negative strains. Both these phases
disappear with increasing $\alpha_{12}$ (the solution obtained in this case
agrees with that of Ref.~\onlinecite{PhysRevB.61.R825}), and the $Cm$ phase
does not appear at all. The optimal value of the $\alpha_{12}$ coefficient for
which the predictions of the phenomenological model are closest to the
experiment was found in Ref.~\onlinecite{ApplPhysLett.96.232902}.
In the phase-field model~\cite{JApplPhys.108.084113} at $T \to 0$ the $Ima2$
phase is absent, and it is not clear whether the $Cm$ solution at high strains
can be associated with our $Ima2$(II) phase. Our results also differ from
those of the phenomenological model,~\cite{PhysSolidState.51.1025}
in which its own set of material constants was used. According to
Ref.~\onlinecite{PhysSolidState.51.1025}, when the lattice parameter $a_0$ is
increased, the ground-state structures change as follows:
$P4mm \to I4cm \to I4/mcm \to Ima2 \to Fmm2$(II)${} \to Cm \to Ima2$(II)${} \to Amm2$,
whereas our calculations predict the sequence
$I4cm \to I4/mcm \to Ima2 \to Cm \to Fmm2$(II)${} \to Ima2$(II) and that the
$P4mm$ and $Amm2$ phases never become the ground-state structures. As for the
comparison with the results of first-principles calculations that took into
account both order parameters,~\cite{JApplPhys.100.084104} in the cited work
the existence regions of the $Ima2$ and $Ima2$(II) phases are separated by
an intermediate $Fmm2$(II) phase. In our calculations, the sequence of the
ground-state structures for stretched films differs from the data of
Ref.~\onlinecite{JApplPhys.100.084104} in that the transition between the
$Ima2$ and $Cm$ phases occurs abruptly, and the $Fmm2$(II) phase is the
ground-state structure only in a narrow region separating the $Cm$ and
$Ima2$(II) phases (Fig.~\ref{fig1}).

\subsubsection{Stressed SrTiO$_3$ films}

We now consider the differences between the phase diagrams obtained under the
fixed-stress boundary conditions for two ways of applying stress and the
phase diagrams obtained above under the fixed-strain conditions.

In biaxially compressed film ($p \parallel [110]$), the transition pressures
of the $I4cm$--$I4/mcm$ and $I4/mcm$--$Ima2$ phase transitions are close
to the values of internal stress at the transition points in films grown on
a cubic substrate (with a maximum deviation of 0.3--0.6~kbar). In contrast,
at the biaxial stretching, the phase diagrams differ much stronger [compare
Fig.~\ref{fig1} and Fig.~\ref{fig2}(b)]. This is because under tension, the
$Fmm2$(II), $Cm$, and $Ima2$(II) phases (in which spontaneous strains caused
by polarization and octahedral rotations lie in the film plane) become
energetically more favorable. At low stresses, a configuration in which the
two components of spontaneous strain are perpendicular to each other is the
most favorable, and at high stresses, a configuration in which the spontaneous
strains are parallel to each other and are directed along the [110] pseudocubic
axis is the most favorable. Due to an additional contribution to the
enthalpy proportional to $\epsilon_1 + \epsilon_2 = -p_{[110]} S^{(\rm 2D)}$
(where a ``two-dimensional'' elastic compliance $S^{(\rm 2D)}$ is the sum of
$S_{11} + S_{22} + 2S_{12}$ elastic compliance tensor components in Voigt
notation), the phase with the highest $S^{(\rm 2D)}$ is characterized by the
strongest decrease of the enthalpy with increasing stress, and so it is
asymptotically the most stable. In our case, this is the $Ima2$(II) phase.

Under uniaxial tension ($p \parallel [001]$), the phases in which the spontaneous
strain is directed along the $z$~axis (under tension) or those in which the
spontaneous strain lies in the film plane (under compression) should be
more energetically favorable. However, we should note that the condition
for the occurrence of a phase transition under uniaxial stress differ
significantly from the condition for the case of biaxial stress. For example,
the lattice parameter $a_0$ at which the $I4/mcm$--$I4cm$ phase transition
occurs in SrTiO$_3$ is 7.2785~Bohr for $p \parallel [110]$ and 7.3018~Bohr
for $p \parallel [001]$. This reduces the transition pressure into the $I4cm$
phase and strongly increases the transition pressures into the $Fmm2$(II) and
$Cm$ phases. As an additional contribution to the enthalpy for
$p \parallel [001]$ is proportional to $\epsilon_3 = -p_{[001]} S_{33}$,
in this case asymptotically the most stable phase is that with the highest
$S_{33}$ value. This phase is also the $Ima2$(II) phase, but because
the transition pressure into this phase under uniaxial compression is
sufficiently high ($\sim$85~kbar), this phase does not become the ground
state in the pressure range considered in this work [Fig.~\ref{fig2}(a)].

As we have seen, the transition from the fixed-strain to the fixed-stress
boundary conditions causes a change in the order of phase transitions: they
become the first-order ones, and the width of the corresponding hysteresis
regions depend on the way of applying stress. We attribute the change in
the order of phase transitions to a well-known renormalization of the
coefficients of the fourth-order terms in the Landau expansion produced by
electrostriction, which results in that the second-order phase transitions
can become the first-order ones. The fact that the lattice parameters (as
well as the atomic positions) near the phase transition points are
significantly different for two ways of applying stress enables to explain
why the width of the hysteresis regions near these points depends on how the
stress is applied. It is important to note that the mere fact of appearance of
the first-order phase transitions in SrTiO$_3$ means that the phenomenological
description of its properties should take into account at least the sixth-order
terms in the power series expansion of the thermodynamic potential.

\subsection{The nature of metastability effects}

The most interesting result of this work is the observation of metastable
phases and phase transitions between them in strained and stressed SrTiO$_3$
films. We believe that the appearance of these phenomena is due to competing
instabilities in this material.

Indeed, these phenomena were absent in the first-principles calculations in
which the octahedral rotations were neglected.~\cite{PhysRevB.71.024102,
PhysRevB.72.144101,JapJApplPhys.44.7134}
These calculations predicted the second-order $P4/mmm \to P4mm$ phase transition
in biaxially compressed films and the second-order $P4/mmm \to Amm2$ one in
biaxially tensile films. According to our calculations, in which octahedral
rotations were neglected, at $p = 0$ the most stable phase is
$R3m$,~\cite{PhysSolidState.51.362}  and the biaxial strain transforms it via
an intermediate $Cm$ phase, respectively, to the $P4mm$ phase (at
$a_0 \approx 7.341$~Bohr) or to the $Amm2$ phase (at $a_0 \approx 7.361$~Bohr).%
    \footnote{The appearance of a gap between the $P4mm$ and $Amm2$ polar phases
    is typical of calculations in which the theoretical lattice parameter of
    cubic SrTiO$_3$ was $a_0 = {}$7.27--7.285~Bohr.~\cite{PhysRevB.71.024102,
    PhysRevB.72.144101,JapJApplPhys.44.7134}
    In Ref.~\onlinecite{PhysRevB.49.5828}, where $a_0 = 7.303$~Bohr, the
    lowest-energy phase at $p = 0$ was already the $R3m$ phase. The
    pseudopotentials and calculation technique used in this work give
    a value of $a_0 = 7.3506$~Bohr, which is closest to the experiment
    ($a_0 = 7.379$~Bohr at 300~K). This effect is obviously associated with an
    effective pressure that exists in all LDA calculations (with underestimated
    lattice parameter) and influences the ferroelectric instability.}
No signs of metastability were observed in these calculations.

When both ferroelectric and antiferrodistortive instabilities in SrTiO$_3$ are taken
into account, the metastability appears already under the fixed-strain
boundary conditions. It is observed in the $Fmm2$(II) and $Ima2$(II) phases
at $a_0 = {}$7.39--7.40~Bohr and in the $Pc$ phase, to which an unstable
$Ima2$(II) phase relaxes at $a_0 = {}$7.32--7.34~Bohr upon condensation of two
unstable phonons at the center and at the boundary of the Brillouin zone (the
energy of the $Pc$ phase is significantly higher than that of the ground-state
$I4/mcm$ or $Ima2$ phases, Fig.~\ref{fig1}). The signs of metastability can
also be observed in the region $a_0 = {}$7.35--7.37~Bohr where \emph{two
different solutions} with the $Cm$ symmetry and different orientation of the
polarization, which both satisfy the stability criterion, appear as a result
of relaxation of unstable $Ima2$(II) and $Fmm2$(II) phases.

A particularly large number of metastable phases appears in SrTiO$_3$ under
the fixed-stress boundary conditions. Under these conditions, the majority of
phases, which were simply unstable under fixed-strain conditions, become
metastable (in Fig.~\ref{fig2} the points corresponding to these phases are
connected by solid lines). An analysis shows that the octahedral rotation
patterns in metastable phases are strongly different from those
in the ground states. For example, for the $Ima2$ phase, which is metastable
under tensile stress, the transition to the ground state requires to change
the direction of the octahedral rotation axis from [001] to [110], and for
the $Pc$ and $Fmmm$ phases, which are metastable under compressive stress,
the transition requires to change the direction from [11$x$] to [001] and from
[010] to [001], respectively. This suggests that each polar phase has its
optimal octahedral rotation pattern which stabilizes this phase and creates
a potential barrier preventing this phase to be easily transformed into another
structure. Such a transformation needs to simultaneously shift a large number of
the oxygen atoms involved in the octahedral rotations and to change the lattice
parameters, so that the appearance of metastability is closely related to the
antiferrodistortive instability. The height of the potential barrier separating
the $Ima2$ and $Ima2$(II) phases (which can be transformed into each other via
an intermediate $Amm2$ phase) is $\sim$24~meV per formula unit at $p_{[110]} = -11$~kbar.
The weakening of the metastability effects in films grown on cubic substrates,
in which spontaneous strain cannot be fully realized, confirms this explanation.
Nevertheless, the competition of the local distortions accompanying the appearance
of polarization and octahedral rotations can be observed even in epitaxial
films.

It should be noted that the metastability effects revealed in SrTiO$_3$ films
are qualitatively different from the effects typical for first-order phase
transitions. In the latter case, the metastability appears in the hysteresis
region where two phases described by different values of the same order
parameter fulfill the stability criterion and so can coexist. In SrTiO$_3$,
the metastability effects are due to a complex interaction between the
polarization and the rotational order parameter via a striction mechanism.
This explains why the obtained solutions turn out to be very sensitive to the
mechanical boundary conditions.

The consequence of the metastability effects in systems with competing
instabilities is the impossibility, in certain situations, to establish whether
the obtained solution is the ground-state structure of a system, or just a
metastable state. As shown in this work, the approach used in most
first-principles investigations of phase transitions (according to which the
ground state is searched by a successive approaching to the structure that
satisfies the stability criterion) often results only in the metastable states.
The ground state in these systems can be found only after exploring the entire
phase diagram and all transitions between the phases. This analysis, however,
can be facilitated by studying the phase diagrams under the fixed-strain
boundary conditions.

It can be seen that the existence of metastable phenomena in SrTiO$_3$ results
in the appearance of bistability regions on the phase diagrams, in which stable
and one or more metastable polar phases coexist. Let us discuss one interesting
possibility of using this bistability. The phase sequence obtained on the
$I4cm$--$I4/mcm$--$Ima2$ branch in stretched SrTiO$_3$ films
[Fig.~\ref{fig2}(b)] enables to offer one more possible application of such
films in non-volatile phase change memory devices. Indeed, by applying a
specific sequence of stresses when cooling the film, a homogeneous metastable state
in the hysteresis region (the $Ima2$ phase) can be prepared in it. Then, the
regions of equilibrium [$Fmm2$(II), $Cm$] or the other metastable [$Ima2$(II)]
phases can be formed on the film surface by the local optical heating. Due to
the bistability of the system, both phases can coexist at low temperatures for
a long enough time, and a contrast in their optical properties can be used for
nondestructive read-out of recorded information. Erasing of information can be
easily realized by heating the film at zero stress.

The sequences of the ground states in SrTiO$_3$ films obtained in this work as
a function of applied strain and stress correspond to the case of $T \to 0$.
At higher temperatures, the phase diagrams may become more complex because of
the appearance of domain-like structures. Such a structure, which consisted of
thick 90$^\circ$ domain walls, was revealed in a first-principles-based effective
Hamiltonian molecular dynamics study of BaTiO$_3$ films subjected to tensile
strain in a wide temperature range below the Curie
temperature.~\cite{ApplPhysLett.107.102901}  In unstrained BaTiO$_3$, the
existence region of this multidomain state coincided with that of the
orthorhombic phase in bulk BaTiO$_3$. However, it remains unclear whether this
multidomain structure is a real (static) one or just a correlated thermal motion
of atoms because the presented data were averaged over a very short (4~ps) time
interval. The existence of competing instabilities in SrTiO$_3$ can complicate
its domain structure. Its complex dynamic behavior and the metastability effects
revealed in this work can result in enhanced loss tangent usually observed in
ferroelectric phases because they both exhibit a retarded reaction to an applied
electric field.

\section{Conclusions}

The first-principles calculations of the phase diagrams of strained and stressed
SrTiO$_3$ films have revealed a number of previously unknown metastability
effects which manifest themselves as coexistence of several phases all satisfying
the stability criterion. These effects become particularly noticeable under the fixed-stress
boundary conditions. It was shown that the cause of these effects is the existence
of competing instabilities in this material. This suggests that similar phenomena
may occur in other ferroelectrics with competing instabilities such as NaNbO$_3$
and BiFeO$_3$.

The metastability effects are also important for understanding the phenomenon of
antiferroelectricity because competing instabilities are very typical for
antiferroelectric materials such as PbZrO$_3$. In particular, the
electric-field-induced polar $R3c$ phase in PbZrO$_3$ is in fact a metastable
phase, in which the octahedral rotation pattern differs from that in the
ground-state nonpolar $Pbam$ structure.

As for the results specific for SrTiO$_3$, it is clear that the phenomenological
description of its properties needs to take into account the higher-order terms
in the Landau expansion, and so the results of previous studies within this
approach require a further examination. \\

\begin{acknowledgments}
The work was supported by Russian Foundation for Basic Research grant No. 13-02-00724.
\end{acknowledgments}

\appendix*
\section{Supplemental material}

\renewcommand{\thetable}{A\arabic{table}}
\setcounter{table}{0}
\setcounter{topnumber}{8}
\setcounter{bottomnumber}{8}
\setcounter{totalnumber}{8}
\setcounter{dbltopnumber}{8}

The energies and enthalpies of different phases calculated for a given strain or stress are compared
in Tables~A1 and A2. The optimized lattice parameters and atomic positions for all structures
considered in this work are given in Tables A3 to A12.

\begin{table*}
\caption{Energies of different phases in strained SrTiO$_3$ film fixed on a square
substrate with a lattice parameter $a_0$ (in meV). The energy of high-symmetry
$P4/mmm$ phase is taken as the energy reference. The energies of stable phases
are in boldface.}
\begin{ruledtabular}
\begin{tabular}{ccccccccc}
$a_0$  & \multicolumn{8}{c}{Phases} \\
(Bohr) & $I4/mcm$      & $I4cm$         & $Fmmm$   & $Ima2$         & $Fmm2$   & $Ima2$(II)     & $Fmm2$(II)     & $Cm$ \\
\hline
7.16  & $-$93.03       & {\bf $-$95.72} & ---      & ---            & ---      & ---            & ---            & --- \\
7.20  & $-$73.46       & {\bf $-$74.53} & ---      & ---            & ---      & ---            & ---            & --- \\
7.22  & $-$64.97       & {\bf $-$65.53} & ---      & ---            & ---      & ---            & ---            & --- \\
7.24  & $-$57.19       & {\bf $-$57.47} & ---      & ---            & ---      & ---            & ---            & --- \\
7.26  & $-$50.17       & {\bf $-$50.28} & ---      & ---            & ---      & ---            & ---            & --- \\
7.27  & $-$46.95       & {\bf $-$46.97} & ---      & ---            & ---      & ---            & ---            & --- \\
7.275 & $-$45.40       & {\bf $-$45.43} & ---      & ---            & ---      & ---            & ---            & --- \\
7.28  & {\bf $-$43.91} & ---            & ---      & ---            & ---      & ---            & ---            & --- \\
7.29  & {\bf $-$41.01} & ---            & ---      & ---            & ---      & ---            & ---            & --- \\
7.30  & {\bf $-$38.28} & ---            & ---      & ---            & ---      & ---            & ---            & --- \\
7.31  & {\bf $-$35.70} & ---            & ---      & ---            & ---      & ---            & ---            & --- \\
7.315 & {\bf $-$34.45} & ---            & ---      & ---            & ---      & ---            & ---            & --- \\
7.32  & $-$33.24       & ---            & $-$26.67 & {\bf $-$33.26} & $-$33.23 & $-$27.67       & ---            & $-$28.21 \\
7.33  & $-$30.94       & ---            & $-$26.89 & {\bf $-$31.07} & $-$30.99 & $-$27.81       & ---            & $-$28.03 \\
7.34  & $-$28.76       & ---            & $-$27.12 & {\bf $-$29.09} & $-$28.96 & $-$27.96       & $-$27.40       & $-$28.00 \\
7.345 & $-$27.71       & ---            & $-$27.24 & {\bf $-$28.19} & $-$28.01 & $-$28.03       & $-$27.63       & $-$28.05 \\
7.35  & $-$26.69       & ---            & $-$27.36 & $-$27.37       & $-$27.12 & {\bf $-$28.17} & $-$27.87       & {\bf $-$28.17} \\
7.36  & $-$24.71       & ---            & $-$27.61 & $-$25.89       & $-$25.47 & $-$28.48       & $-$28.39       & {\bf $-$28.49} \\
7.37  & $-$22.86       & ---            & $-$27.86 & $-$24.65       & $-$24.04 & $-$28.97       & $-$28.99       & {\bf $-$29.04} \\
7.38  & $-$20.98       & ---            & $-$28.02 & $-$23.49       & $-$22.65 & $-$29.44       & $-$29.51       & {\bf $-$29.54} \\
7.39  & $-$19.38       & ---            & $-$28.30 & $-$22.77       & $-$21.62 & $-$30.25       & {\bf $-$30.33} & --- \\
7.40  & $-$17.87       & ---            & $-$28.58 & $-$22.22       & $-$20.77 & $-$31.21       & {\bf $-$31.21} & --- \\
7.42  & $-$15.05       & ---            & $-$29.07 & $-$21.91       & $-$19.61 & {\bf $-$33.55} & $-$33.19       & --- \\
7.44  & $-$12.59       & ---            & $-$29.56 & $-$22.53       & $-$19.29 & {\bf $-$36.65} & $-$35.56       & --- \\
\end{tabular}
\end{ruledtabular}
\end{table*}

\begin{table*}
\caption{Entahlpies of different phases in [110]-stressed SrTiO$_3$ film (in meV). The
energy of high-symmetry $P4/mmm$ phase is taken as the energy reference. The enthalpies of
stable phases are in boldface.}
\begin{ruledtabular}
\begin{tabular}{ccccccccc}
$p_{[110]}$    & \multicolumn{8}{c}{Phases} \\
(kbar)   & $I4cm$         & $I4/mcm$       & $Ima2$   & $Ima2$(II)     & $Imma$   & $Fmmm$         & $Fmm2$(II)     & $Cm$ \\
\hline
44.13    & ---            & ---            & $-$23.08 & {\bf $-$40.48} & ---      & $-$32.40       & $-$36.86       & --- \\
36.78    & ---            & ---            & $-$22.09 & {\bf $-$36.67} & ---      & $-$32.15       & $-$34.88       & --- \\
29.42    & ---            & $-$19.36       & $-$22.21 & {\bf $-$33.86} & ---      & $-$31.96       & $-$33.42       & $-$33.80 \\
22.07    & ---            & $-$21.95       & $-$23.26 & $-$31.79       & ---      & $-$31.70       & $-$32.33       & {\bf $-$32.41} \\
18.39    & ---            & $-$23.28       & $-$24.11 & $-$30.98       & ---      & $-$31.53       & $-$31.90       & {\bf $-$31.91} \\
14.71    & ---            & $-$24.72       & $-$25.15 & $-$30.34       & ---      & $-$31.39       & {\bf $-$31.55\footnotemark[1]} & $-$31.54 \\
11.03    & ---            & $-$26.30       & $-$26.48 & $-$29.95       & ---      & $-$31.39       & {\bf $-$31.45} & --- \\
7.36     & ---            & $-$27.76       & $-$27.81 & $-$29.53       & $-$29.11 & $-$31.21       & {\bf $-$31.24} & --- \\
5.52     & ---            & $-$28.52       & $-$28.54 & $-$29.37       & $-$29.07 & $-$31.10       & {\bf $-$31.12} & --- \\
4.60     & ---            & $-$28.91       & $-$28.91 & $-$29.28       & $-$29.05 & $-$31.06       & {\bf $-$31.08} & --- \\
3.68     & ---            & $-$29.30       & ---      & $-$29.22       & $-$29.02 & {\bf $-$31.02} & ---            & --- \\
0.0      & ---            & {\bf $-$30.88} & ---      & $-$28.97       & $-$28.92 & $-$30.88       & ---            & --- \\
$-$3.68  & ---            & {\bf $-$32.45} & ---      & $-$28.80       & ---      & $-$30.81       & ---            & --- \\
$-$7.36  & ---            & {\bf $-$34.12} & ---      & $-$28.70       & ---      & $-$30.67       & ---            & --- \\
$-$11.03 & ---            & {\bf $-$35.86} & ---      & $-$28.59       & ---      & $-$30.47       & ---            & --- \\
$-$14.71 & {\bf $-$37.69} & $-$37.63       & ---      & $-$28.50       & ---      & $-$30.35       & ---            & --- \\
\end{tabular}
\end{ruledtabular}
\footnotetext[1]{Response function calculations indicate an instability of this phase against its transformation into the $Cm$ phase.} \\
\end{table*}

\begin{table}
\caption{Lattice parameters, Wyckoff positions, and atomic coordinates for the $I4/mcm$ phase
of SrTiO$_3$ at $p = 0$ (the rotations are around the [001] pseudocubic axis).}
\begin{ruledtabular}
\begin{tabular}{ccccc}
\multicolumn{5}{c}{$a = 10.333105$~Bohr, $c = 14.803747$~Bohr} \\
\hline
Atom & WP & $x$ & $y$ & $z$ \\
\hline
Sr & $4b$ & 0.000000 & 0.500000 & 0.250000 \\
Ti & $4c$ & 0.000000 & 0.000000 & 0.000000 \\
O  & $4a$ & 0.000000 & 0.000000 & 0.250000 \\
O  & $8h$ & 0.214346 & 0.714346 & 0.000000 \\
\end{tabular}
\end{ruledtabular}
\end{table}

\begin{table}
\caption{Lattice parameters, Wyckoff positions, and atomic coordinates for the $I4cm$ phase
of SrTiO$_3$ film fixed on a substrate with $a_0 = 7.20$~Bohr (the rotations are around the
[001] pseudocubic axis, the polarization is along the same axis).}
\begin{ruledtabular}
\begin{tabular}{ccccc}
\multicolumn{5}{c}{$a = 10.182338$~Bohr, $c = 14.996956$~Bohr} \\
\hline
Atom & WP & $x$ & $y$ & $z$ \\
\hline
Sr & $4b$ & 0.500000 & 0.000000 & 0.253951 \\
Ti & $4a$ & 0.000000 & 0.000000 & 0.008361 \\
O  & $4a$ & 0.000000 & 0.000000 & 0.249368 \\
O  & $8c$ & 0.207733 & 0.707733 & $-$0.000840 \\
\end{tabular}
\end{ruledtabular}
\end{table}

\begin{table}
\caption{Lattice parameters, Wyckoff positions, and atomic coordinates for the $Fmmm$ phase
of SrTiO$_3$ film fixed on a substrate with $a_0 = 7.38$~Bohr (the rotations are around the
[010] axis).}
\begin{ruledtabular}
\begin{tabular}{ccccc}
\multicolumn{5}{c}{$a = 14.760000$~Bohr, $c = 14.593072$~Bohr} \\
\hline
Atom & WP & $x$ & $y$ & $z$ \\
\hline
Sr & $8h$ & 0.000000 & 0.250507 & 0.000000 \\
Ti & $8d$ & 0.250000 & 0.000000 & 0.250000 \\
O  & $8i$ & 0.000000 & 0.000000 & 0.216530 \\
O  & $8f$ & 0.250000 & 0.250000 & 0.250000 \\
O  & $8g$ & 0.283971 & 0.000000 & 0.000000 \\
\end{tabular}
\end{ruledtabular}
\end{table}

\begin{table}
\caption{Lattice parameters, Wyckoff positions, and atomic coordinates for the $Fmm2$ phase
of SrTiO$_3$ film fixed on a substrate with $a_0 = 7.38$~Bohr (the polarization is along the
[100] pseudocubic axis, the rotations are around the [001] axis; the $F2mm$ setting).}
\begin{ruledtabular}
\begin{tabular}{ccccc}
\multicolumn{5}{c}{$a = 14.760000$~Bohr, $c = 14.703578$~Bohr} \\
\hline
Atom & WP & $x$ & $y$ & $z$ \\
\hline
Sr & $8d$ & 0.006484 & 0.000000 & 0.249962 \\
Ti & $8c$ & 0.258024 & 0.249145 & 0.000000 \\
O  & $8c$ & $-$0.000988 & 0.281480 & 0.000000 \\
O  & $4a$ & 0.218214 & 0.000000 & 0.000000 \\
O  & $4a$ & $-$0.219248 & 0.000000 & 0.000000 \\
O  & $8b$ & 0.246996 & 0.250000 & 0.250000 \\
\end{tabular}
\end{ruledtabular}
\end{table}

\begin{table}
\caption{Lattice parameters, Wyckoff positions, and atomic coordinates for the $Fmm2$(II) phase
of SrTiO$_3$ film fixed on a substrate with $a_0 = 7.38$~Bohr (the polarization is along the
[100] pseudocubic axis, the rotations are around the [010] axis; the $F2mm$ setting).}
\begin{ruledtabular}
\begin{tabular}{ccccc}
\multicolumn{5}{c}{$a = 14.760000$~Bohr, $c = 14.591528$~Bohr} \\
\hline
Atom & WP & $x$ & $y$ & $z$ \\
\hline
Sr & $8c$ & 0.006027 & 0.250548 & 0.000000 \\
Ti & $8d$ & 0.258045 & 0.000000 & 0.250988 \\
O  & $8d$ & 0.499078 & 0.000000 & 0.283937 \\
O  & $8b$ & 0.247904 & 0.250000 & 0.250000 \\
O  & $4a$ & 0.283166 & 0.000000 & 0.000000 \\
O  & $4a$ & $-$0.285274 & 0.000000 & 0.000000 \\
\end{tabular}
\end{ruledtabular}
\end{table}

\begin{table}
\caption{Lattice parameters, Wyckoff positions, and atomic coordinates for the $Ima2$ phase
of SrTiO$_3$ film fixed on a substrate with $a_0 = 7.38$~Bohr (the polarization is along the
[110] pseudocubic axis, the rotations are around the [001] axis; the $I2am$ setting).}
\begin{ruledtabular}
\begin{tabular}{ccccc}
\multicolumn{5}{c}{$a = 10.436896$~Bohr, $c = 14.702085$~Bohr} \\
\hline
Atom & WP & $x$ & $y$ & $z$ \\
\hline
Sr & $4a$ & 0.511582 & 0.000000 & 0.000000 \\
Ti & $4b$ & 0.014611 & 0.001526 & 0.250000 \\
O  & $4a$ & $-$0.004529 & 0.000000 & 0.000000 \\
O  & $4b$ & 0.717532 & 0.218809 & 0.250000 \\
O  & $4b$ & $-$0.219196 & 0.717984 & 0.250000 \\
\end{tabular}
\end{ruledtabular}
\end{table}

\begin{table}
\caption{Lattice parameters, Wyckoff positions, and atomic coordinates for the $Ima2$(II) phase
of SrTiO$_3$ film fixed on a substrate with $a_0 = 7.38$~Bohr (the polarization is along the
[110] pseudocubic axis, the rotations are around the same axis; the $I2am$ setting).}
\begin{ruledtabular}
\begin{tabular}{ccccc}
\multicolumn{5}{c}{$a = 10.436896$~Bohr, $c = 14.594969$~Bohr} \\
\hline
Atom & WP & $x$ & $y$ & $z$ \\
\hline
Sr & $4b$ & 0.510431 & 0.001862 & 0.250000 \\
Ti & $4a$ & 0.012530 & 0.000000 & 0.000000 \\
O  & $4b$ & 0.497252 & 0.547074 & 0.250000 \\
O  & $8c$ & 0.250394 & 0.249830 & 0.024560 \\
\end{tabular}
\end{ruledtabular}
\end{table}

\begin{table}
\caption{Lattice parameters, Wyckoff positions, and atomic coordinates for the $Cm$ phase of
SrTiO$_3$ film fixed on a substrate with $a_0 = 7.36$~Bohr (the polarization and the rotation
axis are in the $xy$ plane; the $Am$ setting).}
\begin{ruledtabular}
\begin{tabular}{ccccc}
\multicolumn{5}{c}{$a = 10.408612$~Bohr, $c = 14.619606$~Bohr\footnotemark[1]} \\
\hline
Atom & WP & $x$ & $y$ & $z$ \\
\hline
Sr & $2a$ & 0.504500 & 0.503322 & 0.000000 \\
Sr & $2a$ & 0.505691 & $-$0.000127 & 0.000000 \\
Ti & $4b$ & 0.005456 & 0.002834 & 0.250500 \\
O  & $2a$ & 0.029421 & 0.454223 & 0.000000 \\
O  & $2a$ & $-$0.029241 & 0.043574 & 0.000000 \\
O  & $4b$ & 0.001801 & 0.248585 & 0.280345 \\
O  & $4b$ & 0.501557 & 0.248085 & 0.734216 \\
\end{tabular}
\end{ruledtabular}
\footnotetext[1]{The translation vectors are [100], [110], and [001].}
\end{table}

\begin{table}
\caption{Lattice parameters, Wyckoff positions, and atomic coordinates for the $Fmm2$(III) phase
of SrTiO$_3$ film subjected to a stress of $p_{[001]} = -14.7$~kbar (the polarization is along
the [001] pseudocubic axis, the rotations are around the [010] axis).}
\begin{ruledtabular}
\begin{tabular}{ccccc}
\multicolumn{5}{c}{$a = 14.59641$~Bohr, $b = 14.77120$~Bohr,} \\
\multicolumn{5}{c}{$c = 14.72473$~Bohr} \\
\hline
Atom & WP & $x$ & $y$ & $z$ \\
\hline
Sr & $8c$ & 0.000000 & 0.249575 & 0.001152 \\
Ti & $8d$ & 0.249152 & 0.000000 & 0.252941 \\
O  & $4a$ & 0.000000 & 0.000000 & 0.210624 \\
O  & $4a$ & 0.000000 & 0.000000 & 0.279767 \\
O  & $8b$ & 0.250000 & 0.250000 & 0.244342 \\
O  & $8d$ & 0.284337 & 0.000000 & $-$0.004654 \\
\end{tabular}
\end{ruledtabular}
\end{table}

\begin{table}
\caption{Lattice parameters, Wyckoff positions, and atomic coordinates for the $Pc$ phase of
[001]-stressed SrTiO$_3$ film ($p_{[001]} = -14.7$~kbar); the $Pa$ setting.}
\begin{ruledtabular}
\begin{tabular}{ccccc}
\multicolumn{5}{c}{$a = 10.351368$~Bohr, $b = 10.405403$~Bohr,} \\
\multicolumn{5}{c}{$c = 14.747770$~Bohr, $\beta = 89.9952^\circ$} \\
\hline
Atom & WP & $x$ & $y$ & $z$ \\
\hline
Sr & $2a$ & 0.507791 & 0.252513 & 0.255092 \\
Sr & $2a$ & 0.497563 & 0.247451 & 0.755063 \\
Ti & $2a$ & 0.002976 & 0.248822 & 0.007148 \\
Ti & $2a$ & 0.002331 & 0.251203 & 0.507150 \\
O  & $2a$ & 0.001328 & 0.204662 & 0.249255 \\
O  & $2a$ & 0.004428 & 0.295355 & 0.749257 \\
O  & $2a$ & 0.244962 & $-$0.007996 & $-$0.024935 \\
O  & $2a$ & 0.760337 & 0.507554 & 0.475052 \\
O  & $2a$ & $-$0.738300 & 0.491565 & 0.022296 \\
O  & $2a$ & $-$0.255419 & 0.007896 & 0.522307 \\
\end{tabular}
\end{ruledtabular}
\end{table}

\clearpage


\providecommand{\BIBYu}{Yu}

\end{document}